*Research article*

# Ecosystem-level stabilizing effects of biodiversity via nutrient-diversity feedbacks in multitrophic systems

**Authors:**

Chun-Wei Chang[a], orcid.org/0000-0002-9817-2956, Email: picachueco@gmail.com

Chih-hao Hsieh[a,b,c,d], orcid.org/0000-0001-5935-7272, Email: chsieh@ntu.edu.tw

Takeshi Miki[b,e,1], orcid.org/0000-0002-2452-8681, Email: tks.miki.ecology@gmail.com

**Author Affiliation:**

[a]Research Center for Environmental Changes, Academia Sinica, Taipei 11529, Taiwan

[b]Institute of Oceanography, National Taiwan University, No. 1, Sec. 4, Roosevelt Rd., Taipei 10617, Taiwan

[c]Institute of Ecology and Evolutionary Biology, Department of Life Sciences, National Taiwan University, Taipei 10617, Taiwan

[d]National Center for Theoretical Sciences, Taipei 10617, Taiwan

[e]Department of Environmental Solution Technology, Faculty of Science and Technology, Ryukoku University, 1-5 Yokotani,Seta Oe-cho,Otsu,Shiga 520-2194. Japan

**Corresponding Author:**

Takeshi Miki
Department of Environmental Solution Technology, Faculty of Science and Technology, Ryukoku University, 1-5 Yokotani,Seta Oe-cho,Otsu,Shiga 520-2194.0020Japan,
tks.miki.ecology@gmail.com
Tel: +81-(0)77-544-7111, Fax: NA

**Abstract**

Statistical averaging and asynchronous population dynamics as portfolio mechanisms are considered as the most important processes with which biodiversity contributes to ecosystem stability. However, portfolio theories usually regard biodiversity as a fixed property, but overlook the dynamics of biodiversity altered by other ecosystem components. Here, we proposed a new mechanistic food chain model with nutrient-diversity feedback to investigate how dynamics of phytoplankton species diversity determines ecosystem stability. Our model focuses on nutrient, community biomass of phytoplankton and zooplankton, and phytoplankton species richness. The model assumes diversity effects of phytoplankton on trophic interaction strength along plankton food chain: phytoplankton diversity influences nutrient uptake by phytoplankton and zooplankton grazing on phytoplankton, which subsequently affects nutrient level and community biomass of phytoplankton and zooplankton. The nutrient level in turn affects phytoplankton diversity. These processes collectively form feedbacks between phytoplankton diversity and dynamics of plankton and nutrient. More importantly, nutrient-diversity feedback introduced additional temporal variabilities in community biomass, which apparently implies a destabilizing effect of phytoplankton diversity on ecosystem. However, the variabilities made ecosystems more robust against extinction of plankton because increasing phytoplankton diversity facilitates resource consumptions when consumers prone to extinct; while, reducing diversity weakens destabilizing dynamics caused by over-growth. Our results suggest the presence of a novel stabilizing effect of biodiversity acting through nutrient-diversity feedback, being independent of portfolio mechanisms.

## Introduction

Biodiversity has been shown as one of the most important determinants affecting ecosystem stability and sustainability (Tilman *et al.* 2014). Stabilizing effect of biodiversity is believed to operate through portfolio mechanisms (Tilman *et al.* 1998; Lehman & Tilman 2000), including statistical averaging (Doak *et al.* 1998) and asynchronous dynamics among individual populations (Yachi & Loreau 1999; McCann 2000). These portfolio mechanisms predict positive relationships between the number of populations and temporal stability of the summed population properties, such as total biomass (Thibaut & Connolly 2013).

Stabilizing effect of portfolio mechanisms is effective only when some strict assumptions on individual populations are fulfilled. For example, stabilization through statistical averaging effects requires a strict mean-variance relationship (Tilman *et al.* 1998) and evenly distributed abundance (Doak *et al.* 1998); stabilization through asynchronous dynamics requires the existence of differential responses to environmental fluctuations among populations (Yachi & Loreau 1999; Loreau & de Mazancourt 2008), i.e., response diversity (Elmqvist *et al.* 2003). Without these population-level assumptions, species diversity might have no effect or even destabilizing effect on temporal stability of community biomass (Thibaut & Connolly 2013).

Apart from the population-level assumptions, portfolio theories neglect the regulatory role of diversity on multitrophic interactions. For instance, plant diversity has been shown to exhibit a positive relationship with nutrient uptake and plant community biomass (Loreau 1998; Cardinale 2011) and a positive or negative relationship with herbivory and herbivore community biomass (Thébault & Loreau 2003; Hillebrand & Cardinale 2004). Hereafter, we call these relationships *multitrophic biodiversity-ecosystem function* (*multitrophic BDEF*). Multitrophic BDEF is potentially an important determinant of stability because strength of multitrophic interactions, such as top-down or bottom-up interactions, critically affects stability of resource-consumer dynamics (McCann *et al.* 1998). Nevertheless, multitrophic BDEF effects on ecosystem stability remain lack of theoretical understanding.

Furthermore, the portfolio theories assume species diversity as a fixed, given property in ecosystems (Lehman & Tilman 2000; de Mazancourt *et al.* 2013) or as an emergent property in metacommunity dynamics (Loreau *et al.* 2003). In either case, species diversity is assumed having no dynamics, and the effects of diversity on ecosystem functioning cannot in turn feedback to changing diversity. These assumptions apparently deviate from empirical observations indicating that species diversity is not a static external driver (Tsai *et al.* 2014) but has mutual feedbacks with its environments and ecosystem functioning (Chapin III *et al.* 2000; Schmid 2002b; Miki 2008). In fact, the existence of diversity-mediated feedbacks has been demonstrated in a recent empirical study which constructs interaction networks among diversity, productivity, and environments (Chang et al. under review). This study points out that the diversity-mediated feedbacks are the consequence of bidirectional interactions between diversity and environments factors (e.g., resources). Firstly, the diversity of primary producers promotes the efficiency of resource consumption, e.g., high phytoplankton diversity promotes the efficiency of nutrient uptake. Secondly, nutrient dynamics as results of phytoplankton consumptions can feedback to the diversity (Chang et al. under review) because nutrient availability is a key determinant of species coexistence (Hsu *et al.* 1977; Tilman 1982; Interlandi & Kilham 2001). In fact, previous studies have indicated a unimodal relationship between nutrient availability and plant diversity (Schmid 2002a; Miki 2008) (*nutrient-diversity feedback*, hereafter). Actually, existence of feedbacks has been shown affecting stability in a variety of dynamical systems (Charlson *et al.* 1987; Miki *et al.* 2010); however, the effects of diversity-mediated feedbacks on stability of ecosystem dynamics is still overlooked in most of current biodiversity theories.

Here, we propose a new mechanism of diversity effect on stability as *ecosystem-level stabilizing effect*, with which multitrophic BDEF and nutrient-diversity feedback interactively determine ecosystem stability. We hypothesize that these two factors are interdependent because 1) multitrophic BDEF regulates ecosystem stability via altering interaction strengths along food chain, and 2) alters nutrient level, in turn, affecting biodiversity via nutrient-diversity feedback, which reinforces or buffers the stabilizing/destabilizing mechanism. Multitrophic BDEF and nutrient-diversity feedback as well as portfolio mechanisms might emerge from population dynamics of interacting multiple species (Duffy *et al.* 2007; Hautier *et al.* 2009; Thibaut & Connolly 2013). However, to test the hypothesis of the ecosystem-level stabilizing effects independently of portfolio effects at population level, we propose a novel theoretical model that does not consider multispecies population dynamics. Our model represents a plankton system, in which species diversity shows clear fluctuations because of rapid turnovers in plankton communities (Tsai *et al.* 2014). In this mathematical model, a dynamical modelling of phytoplankton species richness based on island biogeography theory (IBT) (25, 26), without considering population dynamics of multiple phytoplankton species, is combined to a model of mechanistic plankton food chain. This model has the minimal complexity to include mutual feedbacks between biodiversity, ecosystem functioning, and local environments. First, as a mechanism of multitrophic BDEF, phytoplankton diversity regulates strengths of multitrophic interactions through adjusting the efficiency of resource consumptions. More specifically, phytoplankton nutrient uptake rate and zooplankton ingestion rate on phytoplankton are functions of phytoplankton diversity. Second, phytoplankton diversity, instead of being a fixed parameter, is modelled as a state variable driven by nutrient dependent extinction, representing nutrient-diversity feedback (Kassen *et al.* 2000; Worm *et al.* 2002). Based on this model, we aim to examine how diversity influences temporal stability of ecosystems through multitrophic BDEF and nutrient-diversity feedback.

## Model

We develop a novel model combining simple plankton food chain and dynamical modelling of species richness. This model includes four state variables, nutrient ($N$), phytoplankton species richness ($R$), phytoplankton biomass ($C$), and zooplankton biomass ($Z$). The model framework includes three main parts (Fig. 1A.), 1) multitrophic plankton food chain between $N$, $C$, and $Z$, i.e., NPZ model (Franks 2002), 2) phytoplankton diversity and its effects on nutrient consumption and zooplankton ingestion, and 3) nutrient feedback affecting on phytoplankton diversity. The linkages among these parts forms a mutual feedback between phytoplankton species richness and the plankton food chain (Fig. 1). On the one hand, phytoplankton diversity influences plankton food chain through changing the efficiency of i) nutrient uptake by phytoplankton and ii) phytoplankton ingestion by zooplankton. Both effects are modelled by power-law functions of phytoplankton species richness as suggested in the previous study (Reich *et al.* 2012), where the power-law exponents denoted as $b_N$ and $b_Z$, respectively, are used to quantify the strength of multitrophic BDEF relationship (Table 1). On the other hand, phytoplankton species richness is modelled as a dynamical state variable instead of fixed ecosystem property. This dynamical modelling of species diversity is an extension of IBT (MacArthur & Wilson 1967) to local communities considering the balance between self-regulatory immigration and nutrient dependent extinction processes (Fig. 1B). The dependency of extinction rate on nutrient, $\theta$, is used to determine the strength of nutrient-diversity feedback on species diversity (see Methods). Finally, we examine how varying strengths of nutrient-diversity feedback ($\theta$) as well as multitrophic BDEF ($b_N$ and $b_Z$) influences the temporal variability of dynamical systems. Details on quantification of stability are presented in Methods.

## Results

First, we examined the effect of multitrophic BDEF, i.e., the strength of phytoplankton diversity effects on nutrient uptake ($b_N$) and zooplankton ingestion ($b_Z$), on temporal stability (measured as CV of phytoplankton biomass) of the model plankton system with mild strength of nutrient-diversity feedback ($\theta$=100). The plankton system exhibited diverse dynamical behaviours, including stable equilibrium under high $b_N$ but low $b_Z$ conditions (i.e., CV of all state variables $\leq 10^{-10}$; blue coloured regions in Figs. 2 and S1), fluctuating dynamics (i.e., CV of any state variable $> 10^{-10}$;

warm coloured regions in Fig. 2) under low $b_N$ but high $b_Z$ conditions, and extinctions (levels of any state variable drops below the extinction threshold, $10^{-10}$; grey coloured regions in Fig. 2) under very high $b_Z$. Concerning the $b_N$ effects, the proportion of parameter sets leading to fluctuating dynamics or extinction conditionally decreased with increasing $b_N$ (Fig. 3A) until all parameter sets reached stable equilibria when $b_N$ >0.6. Besides, increase of $b_N$ also leads to decrease of the maximal CV (Fig. 3B). For the $b_Z$ effects, the plankton system stayed in stable equilibrium when $b_Z$- <-0.15 (Figs. 2 and 3C). When further increasing $b_Z$, there were more and more parameter sets leading to fluctuating dynamics until $b_Z \geq 0.2$, with which some systems started to go extinct because of over-fluctuation. As a consequence, the maximal CV exhibited a unimodal pattern with the peak at $b_Z$=0.21 (Fig. 3D). The influences of varying $b_Z$ and $b_N$ on the averaged nutrient concentration, zooplankton biomass, and phytoplankton diversity and biomass are presented in supplemental materials (Fig. S2).

Second, we found that strength of nutrient-diversity feedback influenced the stability of plankton systems in multifaceted ways. Comparing with the basal model without the nutrient-diversity feedback (i.e., $\theta$=0; Fig. S3), the models with the feedback ($\theta$>0) have higher proportion of parameter sets leading to fluctuating dynamics (Fig. 4A), except for the systems with relatively weak feedbacks ($\theta \leq 50$). Furthermore, models with strong feedback strength ($\theta$>250) exhibit greater volatility (i.e., larger maximal CV of phytoplankton biomass) than the base model ($\theta$>0); however, models with moderate feedback strength ($25<\theta<250$) exhibit lower maximal CV than the base model (Fig. 4B). These results indicate that the existence of feedback although breaks the equilibrium maintained at high $b_N$, it makes the plankton systems less volatile, at least under moderate feedback strength (Fig. 4B and S4). Moreover, when strengthening the nutrient-diversity feedback, the proportion of extinction events monotonically decreases (Fig. 4A). This reduction of extinction risk is more significant when phytoplankton diversity effects on nutrient are weaker (i.e., small $b_N$; Fig. S4C) or when phytoplankton diversity effects on zooplankton ingestion are stronger (i.e., large $b_Z$; Fig. S4G).

## Discussion

Our theory, for the first time, demonstrates that the stabilizing/destabilizing effects of diversity through multitrophic BDEF in dynamical ecosystems. Ecosystems are more stable when $b_N$ becomes more positive, implying that a stronger positive relationship between diversity and nutrient uptake rate helps stabilize the dynamical system. In contrast, ecosystems are more stable when $b_Z$ is more negative, i.e., phytoplankton diversity act as a defence mechanism to resist zooplankton ingestion, perhaps due to proliferations of resistant species (Duffy *et al.* 2007) or dilution of suitable species to specialist consumers (Keesing *et al.* 2010). However, ecosystems are less stable when $b_Z$ is more positive, i.e., phytoplankton diversity acts as a balance diet mechanism to facilitate zooplankton ingestion (Duffy *et al.* 2007). Indeed, balance diet mechanism often leads to strong phytoplankton-zooplankton interaction which has been shown unstable (McCann *et al.* 1998). Thus, our model results indicate when the strength of trophic interactions is weakened by defence mechanisms of phytoplankton, phytoplankton diversity exhibits stabilizing effects. As defence mechanisms of resources are common in natural systems (Agrawal 2011), this stabilizing effect through multitrophic BDEF might be prevalent (Hillebrand & Cardinale 2004; García-Comas *et al.* 2016). However, for this stabilizing effect to operate, nutrient feedback on diversity is needed in order to alleviate the negative impacts caused by strong phytoplankton defenses that otherwise causes zooplankton extinction (Fig. S4). In contrast, when the diversity effect is strengthened by the balance diet mechanism, phytoplankton diversity has destabilizing effects. In summary, our results indicate that the multitrophic BDEF acting in a top-down direction (e.g., phytoplankton diversity effects on nutrient, $b_N$, in our case) are generally a stabilizing mechanism; while, the multitrophic BDEF acting in a bottom-up direction (e.g., phytoplankton diversity effects on zooplankton ingestion, $b_Z$, in our case), could be either a stabilizing or destabilizing mechanism depending on the sign of diversity effects (negative or positive, respectively).

Effects of nutrient-diversity feedback ($\theta$) on ecosystem stability are multifaceted (Fig. 4). On the one hand, increasing $\theta$ interrupts the equilibrium conditions maintained by high $b_N$, which makes the systems exhibiting temporal fluctuations in broader parameter regions (Fig. 4A). This effect of increasing $\theta$ is obvious when comparing with the systems lack nutrient-diversity feedback. When $\theta$=0, the diversity reached equilibrium ($R$=16.69) independent of the strength of diversity effects ($b_N$ and $b_Z$) and the other state variables ($R$, $C$, and $Z$). Apparently, the existence of nutrient-diversity feedback introduces temporal fluctuations in diversity, which in turn causes the variability in the strength of mutitrophic interactions, and eventually interrupts the equilibrium conditions. Nevertheless, intermediate strength of the feedback (25<$\theta$<250) reduced the maximal variability in general cases (Fig. 4B), suggesting that the induced fluctuations are better bounded comparing with the base model ($\theta$=0). On the other hand, despite of introducing variability (Fig. S4A, B, E, and F), nutrient-diversity feedback also provides regulatory effects on ecosystem stability and prevents the system collapse caused by plankton extinctions (Fig. 4 and S5). This is because the feedback-induced variability in species richness is beneficial to buffering against destabilizing ecosystem dynamics. Specifically, we found that ecosystems are destabilized when zooplankton-phytoplankton interactions become too strong in the period of $Z$ overgrowth (Fig. S5).

However, this destabilization in our model system was soon alleviated when phytoplankton species diversity decreased, which weakens the destabilizing interactions through reducing zooplankton ingestion, and eventually attenuates the zooplankton overgrowth. In the opposite situation (Fig. S5), when *Z* was vulnerable to extinction, we found that increasing phytoplankton species richness improved zooplankton ingestion efficiency, and thus reduced extinction risk of zooplankton. In both situations, the regulatory dynamics of diversity helped to set up either upper or lower bounds for the fluctuation of state variable, *Z* (Figs. 4B and S5), and thus prevents system collapse. In this sense, nutrient-diversity feedback promotes persistence of the system through bounding fluctuation of nonpoint attractors (Ives & Carpenter 2007) that moves the lower limit of population density away from zero (Hofbauer & Sigmund 1989; McCann 2000) in this non-equilibrium system. Moreover, this regulatory effect is more significant especially when the system lacks stabilizing factors (e.g., weak phytoplankton diversity effects on nutrient $b_N$), or when the system includes strong destabilizing factors (e.g., strong positive phytoplankton diversity effects on zooplankton ingestion $b_N$).

In addition to inducing the regulatory dynamics of diversity, the nutrient-diversity feedback can also affect ecosystems stability by other minor stabilizing mechanisms, including overyielding of phytoplankton biomass and undermining of the negative diversity effects. Firstly, the improvement of overyielding is shown in the relationships between diversity and biomass- the relationships become more positive with increasing strength of the feedback (Figs. S6). Although the improvement is generally minor, the increase of biomass directly contributes to enhance ecosystem stability by decreasing the magnitude of CV (Thibaut & Connolly 2013). Secondly, when the feedback exists, there is a substantial reduction in average phytoplankton diversity when decreasing $b_Z$ (Fig. S2B). The reduction of average phytoplankton diversity can avoid the extinctions caused by too strong negative effects of phytoplankton diversity on zooplankton ingestion ($b_Z \leq -0.44$ in Fig. S4G and H) which strongly suppresses the efficiency of ingestion, and eventually makes zooplankton not self-sustained.

Our dynamical modelling based on extended of IBT reveals a novel stabilizing mechanism of biodiversity operating at ecosystem-level. To the best of our knowledge, this is the first time that IBT has been used to elaborate biodiversity effects on ecosystem functioning and ecosystem level processes, such as nutrient cycling and trophic interactions (MacArthur & Wilson 1967). Based on the extension of IBT, we also integrate two relevant but theoretically loosely connected ideas, biodiversity effects on ecosystem functioning (BDEF) and biodiversity effects on stability, which were considered to be from distinct mechanisms (e.g., niche complementary (Loreau 1998) and portfolio effects (Tilman *et al.* 2014), respectively). In particular, we demonstrate how ecosystem stability can be affected by strength of multitrophic BDEF which is in turn regulated by the nutrient-diversity feedback in dynamical ecosystems. In addition, the proposed ecosystem-level stabilizing effects also support previous empirical findings that ecosystem stability are statistically associated with the strength of biodiversity effects on ecosystem functioning (BDEF) (Chang et al. under review). Therefore, our theoretical dynamical modelling based on extended IBT successfully reveals the connections between biodiversity effects on ecosystem functioning and ecosystem stability, presents high compatibility to empirical findings, and thus sheds light on future biodiversity studies under the frameworks of dynamical ecosystems.

## **Methods**

## Mathematical model

Our theoretical model includes three main parts (Fig. 1A.): 1) multitrophic plankton food chain between *N*, *C*, and *Z* (Franks 2002), 2) phytoplankton diversity *R* and its effects on nutrient consumption and zooplankton ingestion, and 3) nutrient feedback on regulating phytoplankton diversity. In the first part, plankton food chain is parameterized mainly based on Hasting and Powell's tritrophic food chain model (Hastings & Powell 1991), except for the modelling of nutrient instead of basal species. Besides, we include a quadratic term in phytoplankton biomass growth (Table 1; eqn. 3) to specify the growth suppression caused by light shading or the other self-regulatory controls in addition to nutrient competition.

The second part of the model elaborates diversity effects on resource consumptions. Resource consumptions, including nutrient uptakes by phytoplankton and phytoplankton ingestions by zooplankton, are both modelled by Holling type II response functions. The efficiency of consumptions is modelled by power-law functions of phytoplankton diversity (Reich *et al.* 2012), where the strength of diversity effect depends on the value of the power-law exponent denoted as $b_N$ and $b_Z$, for the diversity effects on nutrient uptake and zooplankton ingestions, respectively. The larger power-law exponent indicates a higher sensitivity of resource consumption relative to diversity changes, i.e., stronger diversity effects. When the power-law exponent is equal to zero, the resource consumption is reduced to simple Holling type II response function independent of diversity. In this study, we numerically solve the model covering a wide range of diversity effects. For the diversity effects on nutrient uptake ($b_N$ in Table 2), we consider only the positive, facilitative effects as suggested in most of experimental studies (Hector *et al.* 1999; Tilman *et al.* 2014) ($b_N \geqq 0$), where the increase of phytoplankton species richness improves the efficiency of nutrient uptake. For the diversity effects on zooplankton ingestion ($b_Z$ in Table 2), we consider both positive ($b_Z$ >0) and negative ($b_Z$ <0) diversity effects, because there is no consensus about which effect is stronger in natural systems. The negative diversity effect may be caused by the defences of phytoplankton to zooplankton grazing (García-Comas *et al.* 2016) and the positive diversity effect may be caused by balance diet to facilitate zooplankton growth (Duffy *et al.* 2007).

The third part of the model is a modification of IBT dynamical modelling species diversity driven by the balance between immigration and nutrient feedback. First, the rate of immigration is modelled as a linear decreasing function of species richness since a community with higher species richness is less likely to allow further immigration of other species (MacArthur & Wilson 1967). Second, the extinction rate is modelled as an increasing function of species richness. However, the extent of increasing extinction rate (i.e., the extinction coefficient, $E_X$ ($N$) defined as the eqn. 2 in Table 1 and Fig. 1C) depends on nutrient concentration, and is modelled by a parabolic concave-up function of nutrient (formula 2 in Table 1). This parabolic concave-up function arises because the extinction risk is higher when nutrient is too low to support many species (Rosenzweig 1995), or when nutrient is so high that facilitates the competition exclusion of minor species by dominant species (Isbell *et al.* 2013). As a consequence, the equilibrium of species richness (the black dots in Fig. 1B) calculated from the eqn. 1 in Table 1 changes with the given nutrient concentration and reaches the maximal value at intermediate nutrient level (i.e., unimodal resource-diversity relationship (Miki 2008)). More importantly, in this nutrient-extinction function, we design a curvature parameter $\theta$ to indicate the dependency of the extinction coefficient on nutrient, where the extinction coefficient changes more steeply with nutrient concentration under large $\theta$. Therefore, $\theta$ can be interpreted as the strength of nutrient feedback on extinction, and thus phytoplankton species richness.

**Numerical analysis of theoretical dynamical modelling**

We numerically solved the ODE model under a wide range of parameter sets ($b_N$, $b_Z$, $\theta$), to investigate the influence of diversity effects on the plankton system. We solved the ODE model under 100*100 sets of parameter combinations ($b_N$, $b_Z$) within the ranges, $0 \leqq b_N \leqq 1.0$ and $-0.5 \leqq b_Z \leqq 1.0$, respectively. We repeated the numerical calculations of the 100*100 parameters sets and varied the strength of nutrient feedback on diversity ($\theta$=0, 25, 50, 75, 100, 125, 150, 175, 200, 250, 300, 350, 400, and 450). We collect the time series from the final 10 years from the 110 years' numerical calculations (numerically solved by 4th-order Runge-Kutta algorithm with a fixed step=0.01 day). For each numerical solving step, we set an extinction threshold equal to $10^{-10}$. Any value of state variables dropped below this threshold will be automatically forced to be zero, which terminates the numerical solution and is counted as an extinction event. Based on our model framework, the temporal variability as a consequence of chaotic or torus dynamics can be observed under many parameters sets. Thus, we quantify the temporal variability as the coefficient of variation (CV) estimated from the derived time series. We also set the detection limit for the variability as $CV \geq 10^{-10}$; otherwise, the system is thought to reach equilibrium state when $CV < 10^{-10}$

**Quantification of temporal variability in ecosystem dynamics**

Responses of temporal variability to various strength of diversity effects on nutrient uptake $b_N$ and zooplankton ingestion $b_Z$, and strength of nutrient feedback on diversity $\theta$, are examined. To evaluate the stabilizing effects, we calculated instability indices, including the maximal CV, denoted as $CVx$, and the proportion of parameter sets leading to unstable fluctuations (CV > the detection limit, $10^{-10}$) or leading to extinction (i.e., any state variable dropped below the extinction threshold, $10^{-10}$, at any time step), denoted as $P_c$. These two instability indices were evaluated with respect to different parameters, $b_N$, $b_Z$, and $\theta$, respectively using a conditional computation. For example, to calculate the conditional instability indices with respect to $b_N$, (denoted as $P_c|b_N$ and $CV_x|b_N$), we firstly fixed $b_N$ to a given value, $b_N$*, and then computed the two indices from all the simulation results under each unique parameter set, ($b_N$=$b_N$*, -0.5 ≦ $b_Z$ ≦ 0.5). For simplicity, we only demonstrate the numerical results under $\theta$=100 in the main texts. Next, we gradually varied the given value $b_N$*, and repeated the calculation of instability indices. As such, we can examine how the instability indices change with varying $b_N$* and build up their quantitative relationships. Similarly, the conditional instability indices with respect to $b_Z$, (denoted as $P_c|b_Z$ and $CV_x|b_Z$) can also be calculated following the same procedure except for carrying out the computations under different parameter set with varying $b_Z$*, (0 ≦ $b_N$ ≦ 1.0, $b_Z$=$b_Z$*, $\theta$=100). Finally, we applied this conditional computation to examine the stabilizing effects of $\theta$. Here, we assigned a given value for $\theta$=$\theta$*, and then calculated the instability indices from all the simulation results under each unique parameter set, (0 ≦ $b_N$ ≦ 1.0, -0.5 ≦ $b_Z$ ≦ 0.5, $\theta$=$\theta$*). By repeating the computations of instability indices with varying $\theta$*, we examine the quantitative relationship between $\theta$ and the conditional instability indices, $P_c|\theta$ and $CV_x|\theta$.

**Author contribution:** C.W.C., C.H.H. and T.M. designed the study. C.W.C. performed the computations with help from C.H.H. and T.M. C.W.C., C.H.H., and T.M. wrote the manuscript. The authors declare no conflict of interest.

**Acknowledgements:** This work was supported by the National Center for Theoretical Sciences, Foundation for the Advancement of Outstanding Scholarship, and the Ministry of Science and Technology, Taiwan. T. M. is partly supported by Sumitomo Foundation a Grant for an Environmental Research Project (2018) and MAGIC project (MOST, Taiwan)

## Figure Legends

Fig. 1. Framework of the coupled diversity-ecosystem functioning model. (*A*) This model consists of four variables, *R*: phytoplankton species richness, *N*: nutrient stock, *C*: phytoplankton biomass, and *Z*: zooplankton biomass. In this model, (*B*) the dynamics of specie richness are driven by a balance between species immigration and extinction rate- the dash line indicates how immigration rate decreases with species richness; the solid lines indicate extinction rate as increasing functions of species richness observed under different nutrient levels ($N$=0.35, 0.45, and 0.6). The dots at the line crosses indicate when immigration rate equals to extinction rate, i.e., the equilibria of species richness if the diversity-extinction relationships are fixed. (*C*) The extent of how extinction rate (i.e., extinction coefficient) changes is described as a concave up function of nutrient concentration. The curvature of this nutrient-extinction function is determined by the parameter $\theta$ that quantifies the strength of nutrient feedback on extinction rate, and in turn, species richness.

Fig. 2. Coefficient of variation (CV) measured from the time series of (*A*) phytoplankton biomass and (*B*) zooplankton biomass changes with the strength of diversity effects on nutrient uptake ($b_N$) and zooplankton ingestion ($b_Z$). The other parameters are fixed throughout mathematical computations (e.g., $\theta$=100) using the default values listed in Table 2. Warmer colours represent higher CV for a given parameter set ($b_N$, $b_Z$); blue colour represents CV< the detection limit, $10^{-10}$ (i.e., stable equilibrium); grey colour represents extinction (any state variable dropped below the extinction threshold).

Fig. 3. Diversity effects of phytoplankton influence the stability of plankton system. With increasing diversity effects on nutrient uptake $b_N$, (*A*) the proportion of parameter sets leading to unstable fluctuation and extinction decreased. Similarly, (*B*) the maximal CV of phytoplankton biomass also decreased with increasing $b_N$. However, when increasing the diversity effects on zooplankton ingestion $b_Z$, (*C*) the proportion of parameter sets leading to unstable fluctuation increases until $b_Z$=0.20. Further increase of $b_Z$ ($b_Z \geq 0.21$) causes fierce fluctuation, leading to some systems crushed. Thus, the (*D*) maximal CV shows a unimodal pattern and peaks at around $b_Z$=0.2. All the analyses in this figure are done using the nutrient feedback strength on phytoplankton diversity, $\theta$=100.

Fig. 4. Strength of nutrient feedback $\theta$ on species diversity affects stability of model systems. (*A*) Comparing with the base model ($\theta$=0), more parameter sets lead to unstable fluctuations but less parameter sets lead to extinction when $\theta$ >0. (*B*) The low maximal CV can be observed up to intermediate $\theta$ (i.e. $\theta$<250); whereas, the maximal CV increases with $\theta$ when $\theta \geq 250$. Each summary statistic in panel *A* and *B* was calculated based on the computation results of 100*100 parameter sets under a given $\theta$.

## Figures in Main Text

**Fig. 1**

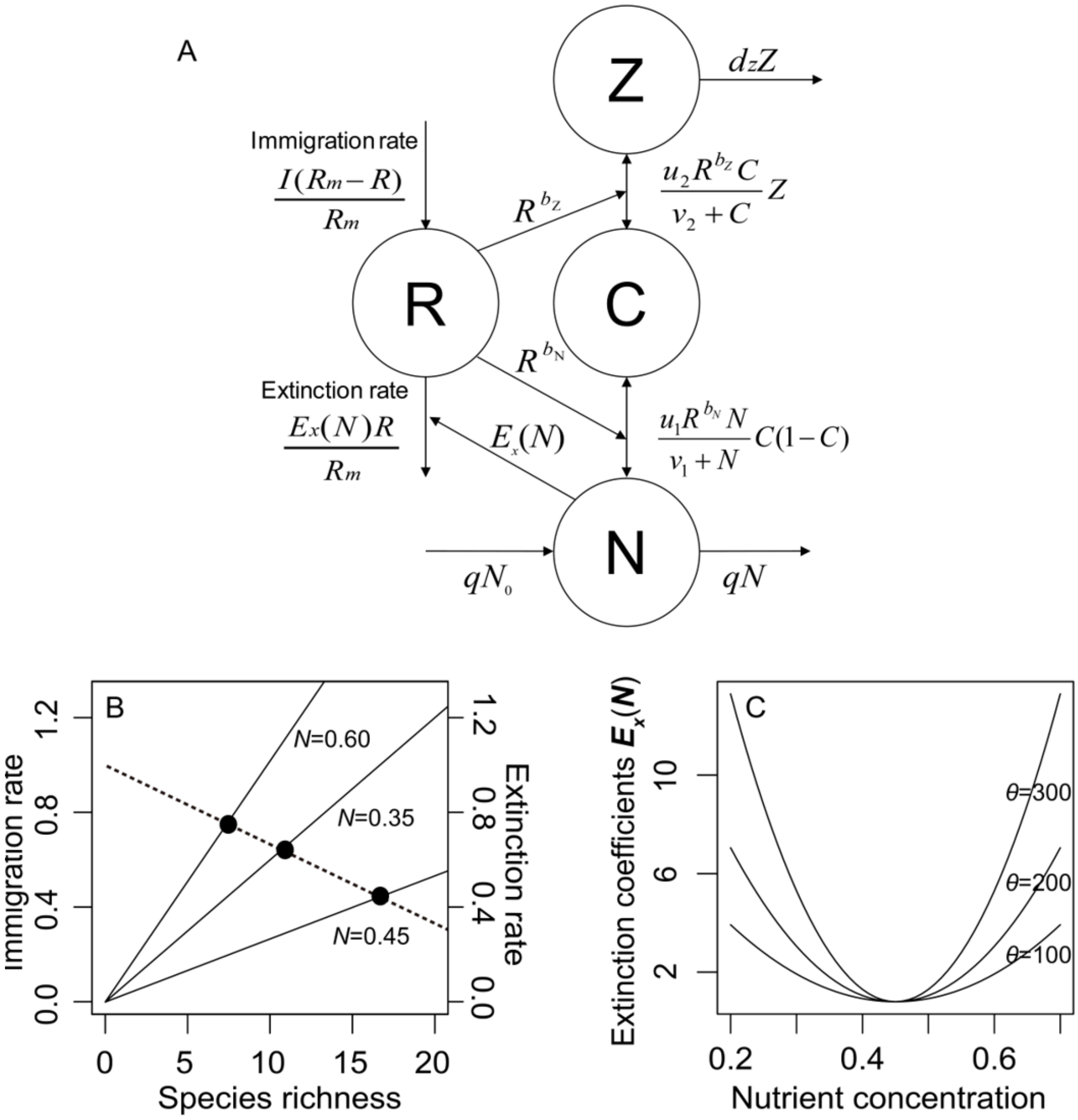

**Fig. 2**

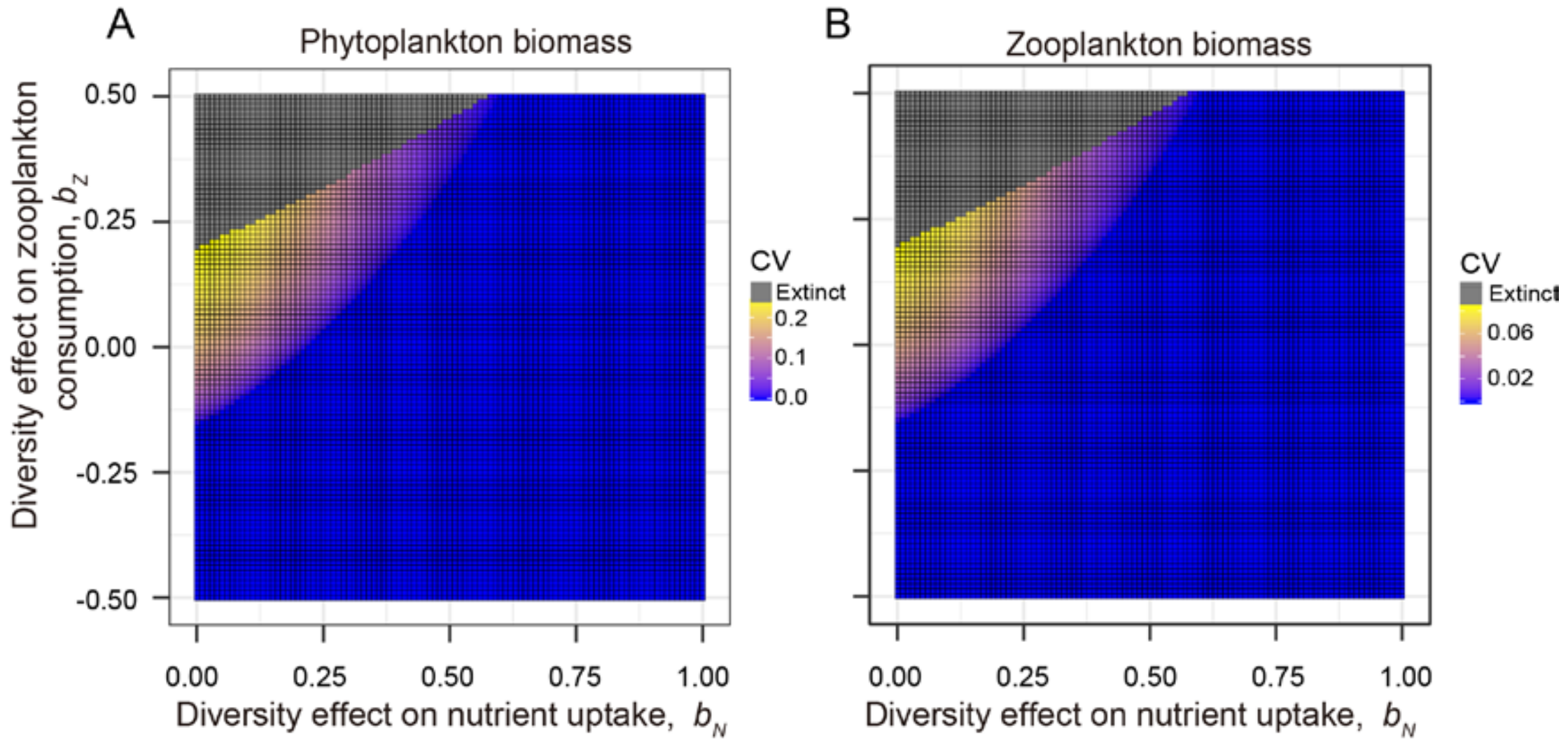

**Fig. 3**

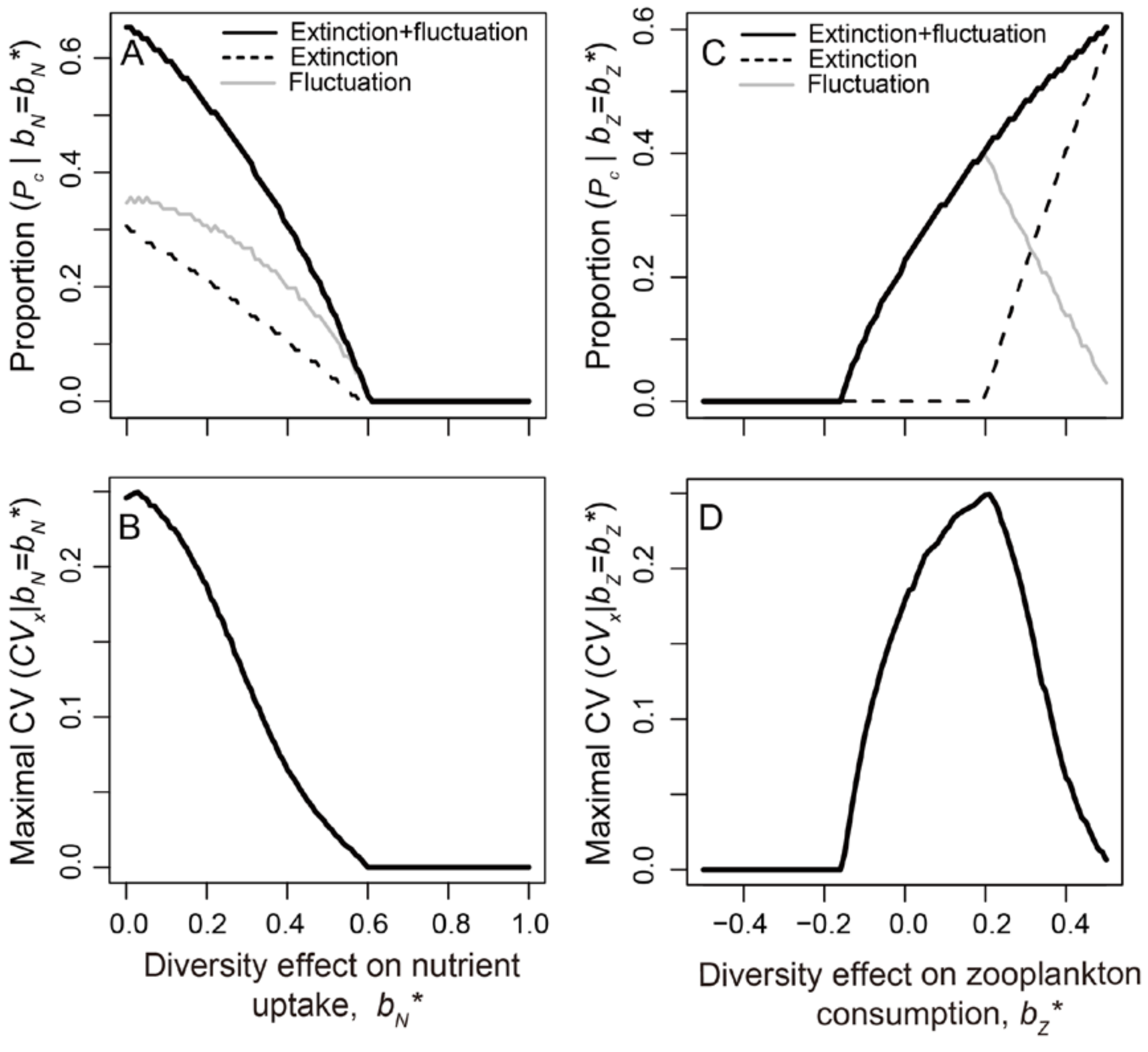
A
B
C
D
Extinction+fluctuation
Extinction
Fluctuation
Proportion ($P_c$ | $b_N$=$b_N$*)
Maximal CV ($CV_x$|$b_N$=$b_N$*)
Proportion ($P_c$ | $b_Z$=$b_Z$*)
Maximal CV ($CV_x$|$b_Z$=$b_Z$*)
Diversity effect on nutrient uptake, $b_N$*
Diversity effect on zooplankton consumption, $b_Z$*

**Fig. 4**

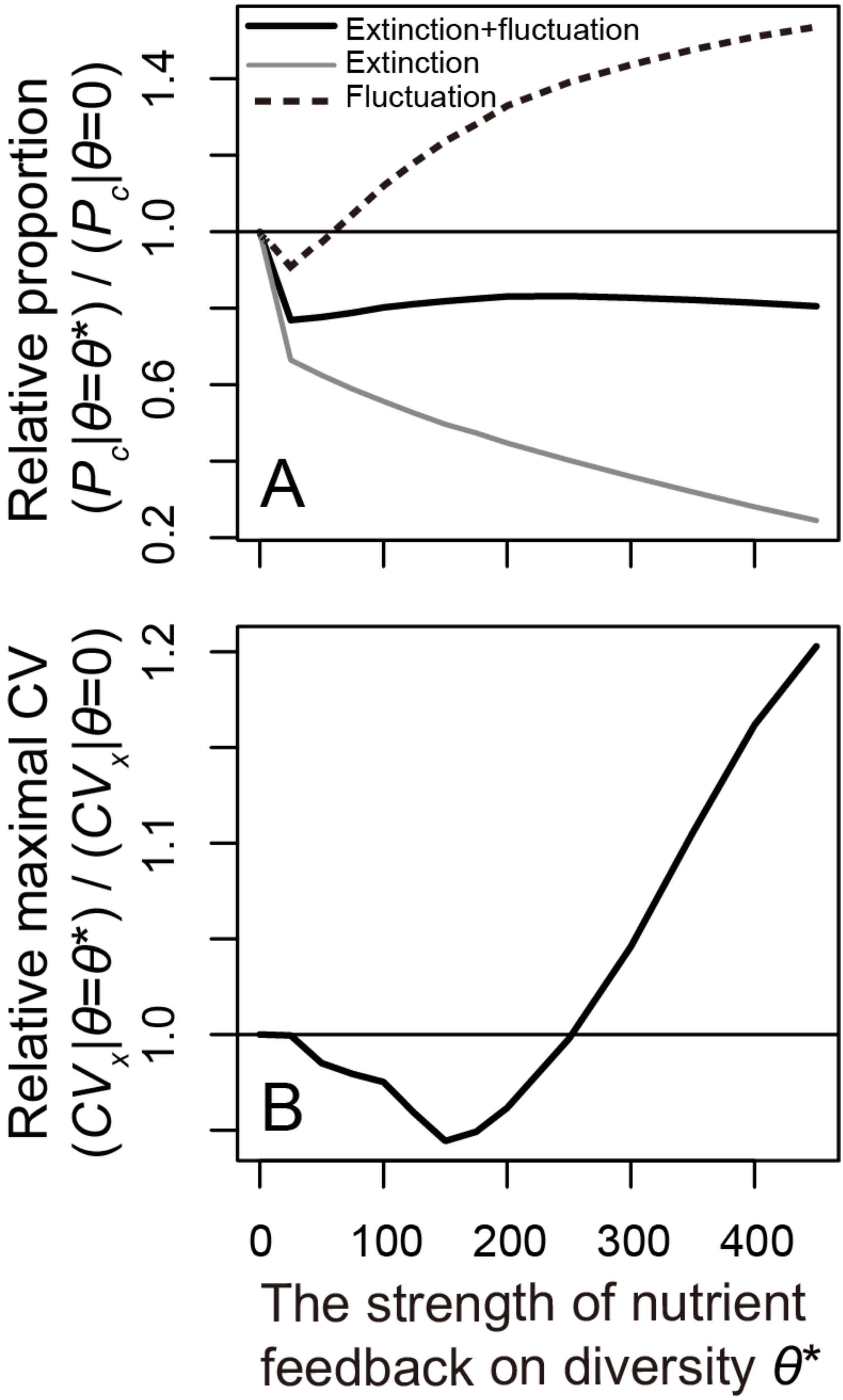

**Table 1.** Model equations and descriptions

| | |
|---|---|
| Phytoplankton species richness (*R*): $\frac{dR}{dt} = \frac{I(R_m - R)}{R_m} - \frac{E_x(N)R}{R_m}$ | (1) |
| = (immigration rate) – (extinction rate with extinction coefficients, $E_X(N) = \theta N^2 - \theta e_b N + e_c(1 + e_b^2(\theta - 1)(4e_c)^{-1})$ | (2)) |
| Phytoplankton biomass (*C*): $\frac{dC}{dt} = u_1 \mathrm{R}^{b_N} \frac{N}{k_1 + N} C(1 - C) - u_2 \mathrm{R}^{b_Z} \frac{C}{k_2 + C} Z$ | (3) |
| = (biomass growth from nutrient uptake) - (biomass reduction due to zooplankton ingestion) | |
| Nutrient concentration (N): $\frac{dN}{dt} = q(N_0 - N) - u_1 \mathrm{R}^{b_N} \frac{N}{k_1 + N} C)$ | (4) |
| = (The nutrient in/out flux caused by system turnover) – (nutrient uptake of phytoplankton) | |
| Zooplankton biomass (*Z*): $\frac{d\mathrm{Z}}{dt} = t_z u_2 \mathrm{R}^{b_Z} \frac{C}{k_2 + C} Z - d_z Z$ | (5) |
| = (biomass growth from consuming phytoplankton) – (natural mortality) | |

**Table 2.** Parameters in dynamical species diversity-ecosystem functioning model

| Symbol | Name | Unit | Default |
|---|---|---|---|
| $b_Z$ | The phytoplankton diversity effect on zooplankton ingestion | Unitless | 0.2 |
| $b_N$ | The phytoplankton diversity effect on the nutrient uptake rate | Unitless | 0.2 |
| $\theta$ | The feedback strength of nutrient on the phytoplankton extinction rate | Unitless | 100 |
| $e_b$, $e_c$ | The constants in the parabolic feedback function, $f_N(N)$ | Unitless | -0.9/1 |
| $u_1$ | Maximum ingestion rate of zooplankton in a single species community | $day^{-1}$ | 0.3 |
| $u_2$ | Maximum nutrient uptake rate of phytoplankton in a single species community | $day^{-1}$ | 0.6 |
| $k_1$ | Half-saturation coefficient of Monod equation, $g_Z$ | $N\ g\ l^{-1}$ | 3.3 |
| $k_2$ | Half-saturation coefficient of Monod equation, $g_N$ | $N\ g\ l^{-1}$ | 2.5 |
| $I$ | The maximal species colonization rate | $day^{-1}$ | 1 |
| $R_m$ | The carrying capacity of species richness | species | 30 |
| $q$ | Nutrient turnover rate | $day^{-1}$ | 0.2 |
| $N_0$ | Inflowing nutrient concentration | $\mu$mol N | 0.55 |
| $t_z$ | Assimilation efficiency of zooplankton ingestion | Unitless | 0.4 |
| $d_Z$ | The natural mortality of zooplankton | $day^{-1}$ | 0.01 |

## Supplementary Information for

# The dynamical stabilizing effects of biodiversity in multitrophic systems

Chun-Wei Chang, Chih-hao Hsieh, Takeshi Miki

Corresponding author: Takeshi Miki

Email: tks.miki.ecology@gmail.com

**This file includes:**

**Fig. S1.**: The coefficient of variation (CV) measured in the time series of phytoplankton species richness and nutrient concentration changed with the strength of diversity effects on nutrient uptake ($b_N$) and zooplankton ingestion ($b_Z$).

**Fig. S2.**: The mean phytoplankton biomass, nutrient concentration, phytoplankton species richness, and zooplankton biomass changed with the strength of diversity effects on nutrient uptake ($b_N$) and zooplankton ingestion ($b_Z$).

**Fig. S3.**: The coefficient of variation (CV) measured in the time series of phytoplankton and zooplankton biomass in base model ($\theta$=0).

**Fig. S4.**: Nutrient feedback on phytoplankton diversity ($\theta$) buffers against the destabilizing dynamics and promotes the persistence of state variables.

**Fig. S5.**: Dynamics of species diversity generated by nutrient-diversity feedback buffers destabilization in ecosystem.

**Fig. S6.**: The slope of BDEF relationship based on the linear regression between average species richness, nutrient, and plankton biomass.

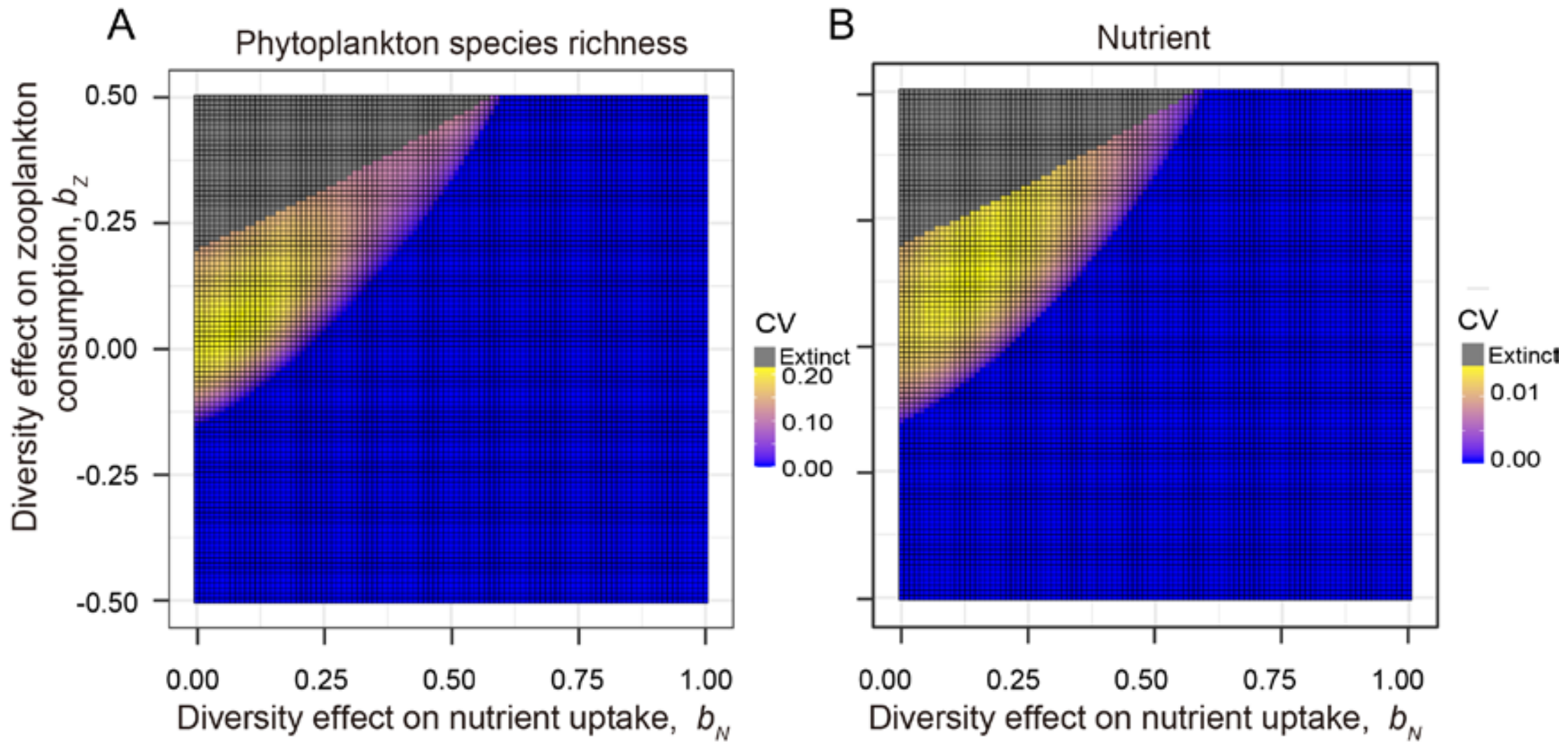

Fig. S1. Coefficient of variation (CV) measured from the time series of (*A*) phytoplankton species richness and (*B*) nutrient concentration changed with the strength of diversity effects on nutrient uptake ($b_N$) and zooplankton ingestion ($b_Z$). Here, all the results are computed based on the default parameter setting (Table 2), which is exactly the same as that used in Fig. 2. Warmer colours represent higher CV measured under a given parameter set ($b_N$, $b_Z$); blue colour represents CV < detection limit, $10^{-10}$ (i.e., stable equilibrium); grey colour represents that extinction happened (any state variable dropped below the extinction threshold).

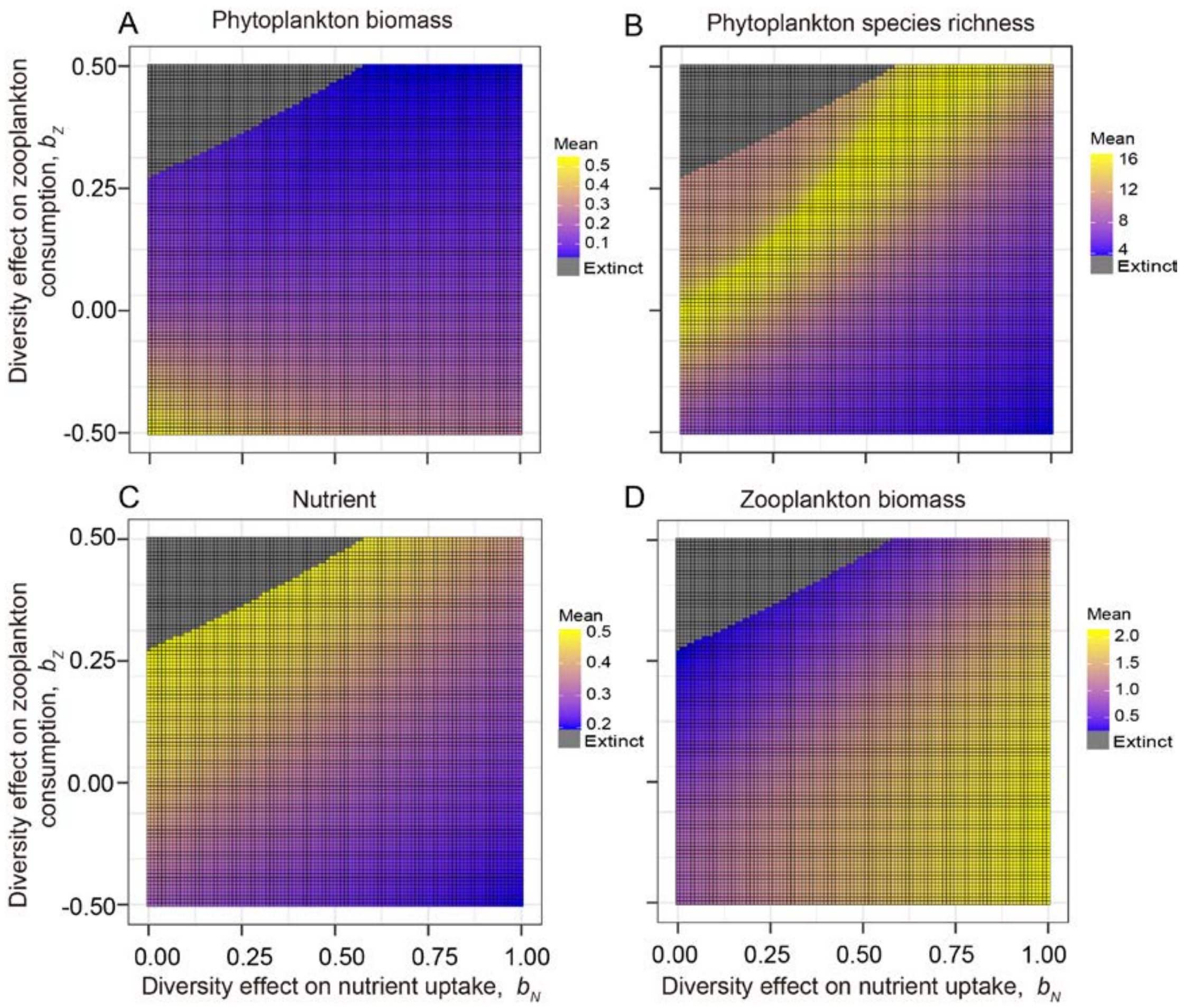

Fig. S2. Mean (*A*) phytoplankton biomass, (*B*) phytoplankton species richness, (*C*) nutrient concentration, and (*D*) zooplankton biomass changed with the strength of diversity effects on nutrient uptake ($b_N$) and zooplankton ingestion ($b_Z$). Here, all the results are computed based on the default parameter setting (Table 2), which is exactly the same as that used in Fig. 2. Warmer colors represent higher mean values measured under a given parameter set ($b_N$, $b_Z$). Grey colour represents that extinction occurred (any state variable dropped below the extinction threshold, $10^{-10}$).

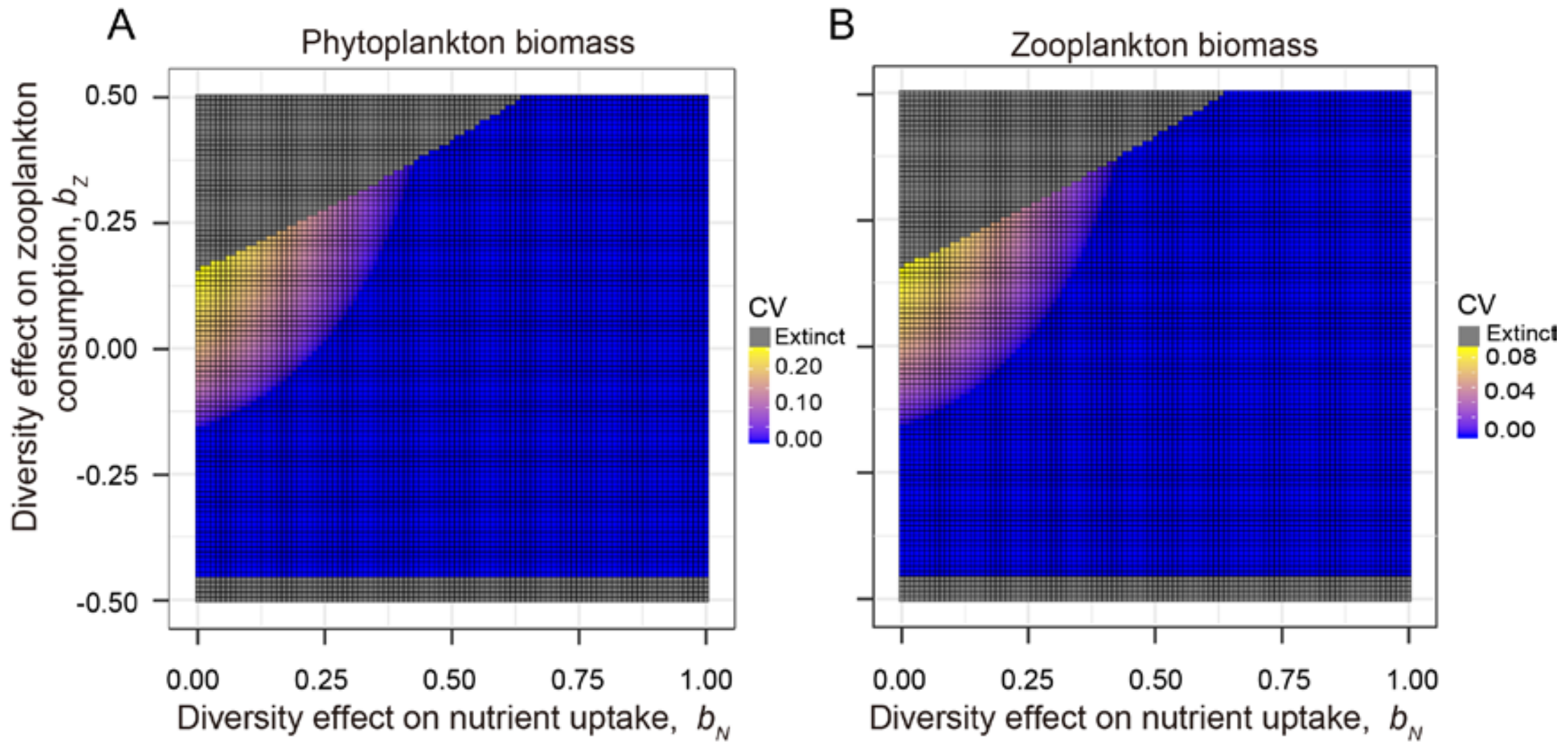

Fig. S3. Coefficient of variation (CV) measured from the time series of (*A*) phytoplankton and (*B*) zooplankton biomass in base model ($\theta$=0) changed with the strength of diversity effects on nutrient uptake ($b_N$) and zooplankton ingestion ($b_Z$). Here, all the results are computed based on the default parameter setting (Fig. 2), except that $\theta$ is changed from 100 to 0. Warm colours represent higher CV measured under a given parameter set ($b_N$ $b_Z$); blue colour represents stable equilibrium (i.e., CV < detection limit, $10^{-10}$); grey colour represents that extinction happened (any state variable dropped below extinction threshold, $10^{-10}$).

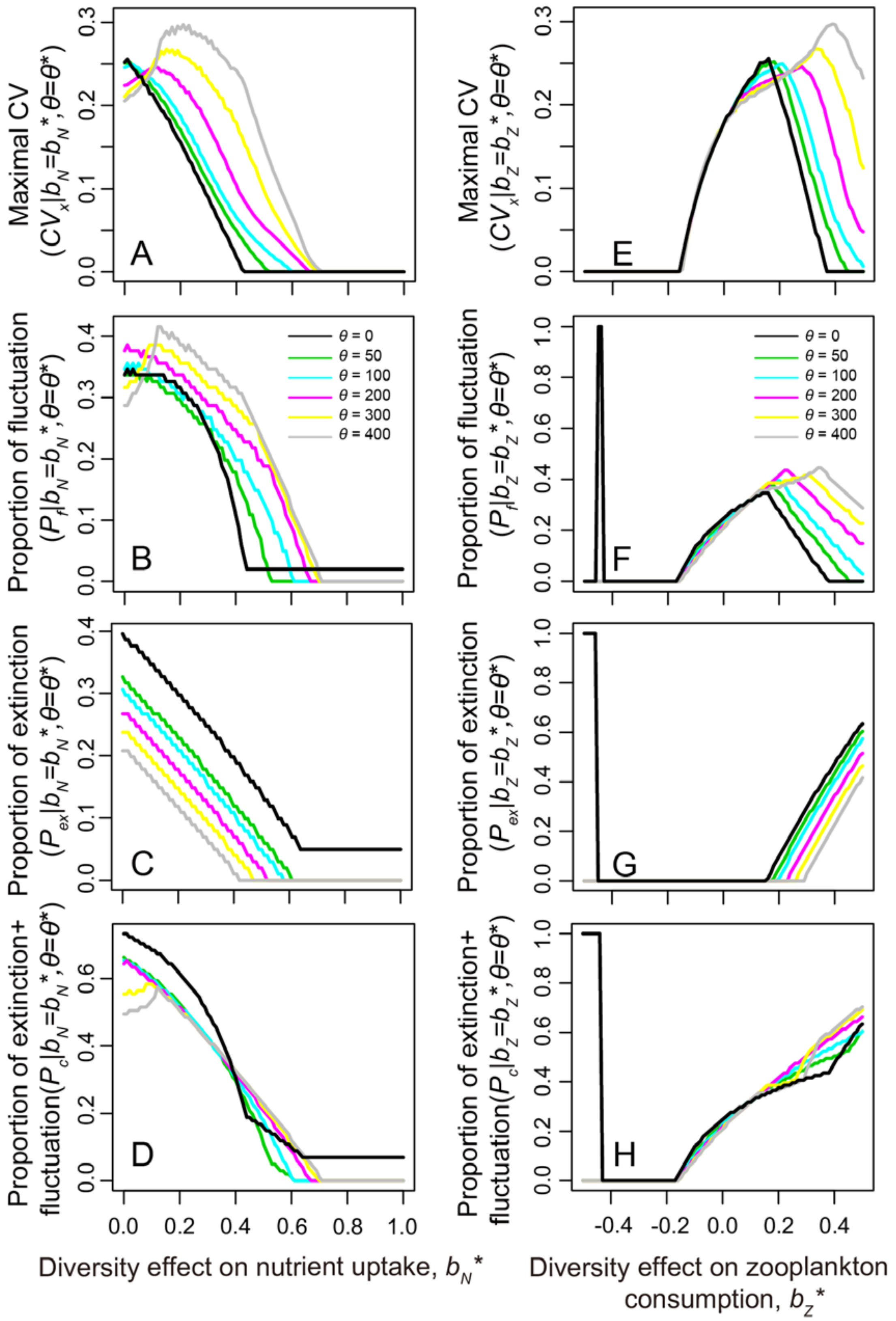

Fig. S4. Strength of nutrient feedback on phytoplankton diversity ($\theta$) and the diversity effects ($b_N$ and $b_Z$) interactively influence the stability of plankton systems. When increasing the diversity effects on nutrient uptake $b_N$, (*A-D*), the maximal CV of phytoplankton biomass decreased (*A*), despite the larger maximal CV is observed in the systems with stronger feedback ($\theta$). Similarly, the proportion of fluctuating systems generally decreased with increasing $b_N$, although a slight increase was observed when $0 \leq b_N \leq 0.2$ and $\theta > 200$ (*B*). The proportion of extinction shows a monotonic decreasing trend regardless of feedback strength; however, the systems with stronger feedback always lead to less extinction outcomes (*C*). Overall, the proportion of unstable systems (fluctuation + extinction) demonstrates a nearly monotonic deceasing trend with increasing $b_N$; while the model with nutrient feedback ($\theta > 0$) usually has lower proportion of unstable systems than the base model ($\theta = 0$) (*D*). In contrast, when increasing diversity effects on zooplankton ingestion $b_Z$, (*E-H*), both the maximal CV (*E*) and the proportion of fluctuated systems (*F*) show unimodal patterns peaking at positive, intermediate levels of $b_Z$. This is because the extinction events were initiated due to over-fluctuation when further strengthening $b_Z$. (*G*). However, when the feedback is stronger, the extinction events are more difficult to be realized (i.e., extinctions happen only when $b_Z$ is very large; *G*). It is worth noting, when lack of the feedback ($\theta = 0$), too strong negative effects of phytoplankton diversity on zooplankton ingestion ($b_Z \leq -0.44$; H) makes system fluctuating ($-0.45 \leq b_Z \leq -0.44$; F) and eventually causes zooplankton to go extinct ($b_Z \leq -0.46$; G). In short, existence of nutrient-diversity feedback ($\theta > 0$) rescues the vulnerable zooplankton populations from extinction under very negative $b_Z$ (panel *H* and Fig. S3), as well as buffers against the extinction events triggered by strengthening $b_Z$ (*G*).

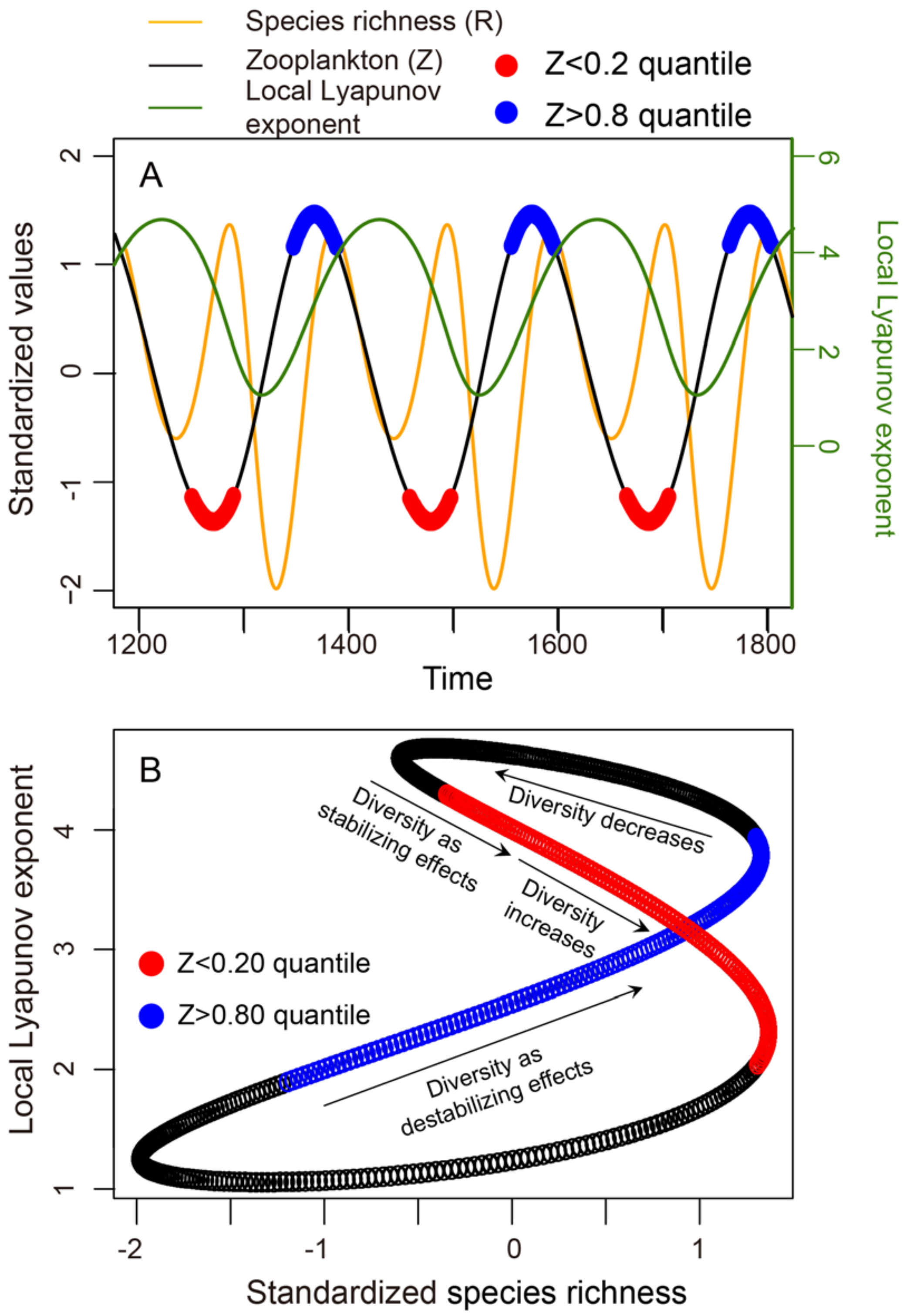

Fig. S5. Dynamics of species diversity generated by nutrient-diversity feedback buffers against system destabilization. (*A*) The example time series of species richness *R*, zooplankton biomass *Z*, and local Lyapunov exponent calculated from the maximal eigenvalue of Jacobian matrix ($b_N$=0.10, $b_Z$=-0.07, $\theta$=100). A smaller local Lyapunov

exponent indicates a more stable ecosystem. The red and blue points labelled the conditions of $Z<0.2$ quantiles and $Z>0.8$ quantiles, respectively. (*B*) When $Z$ is approaching very small values, the increase of species richness helps stabilize the ecosystem dynamics, and thus reduces the probability of zooplankton extinction. When $Z$ grows rapidly, the increase of species richness becomes a destabilizing force to the system; however, the destabilized system soon reduces diversity, which in turn attenuates the overshoot of $Z$ by decreasing resource consumption efficiency, and thus avoids further destabilization of the system.

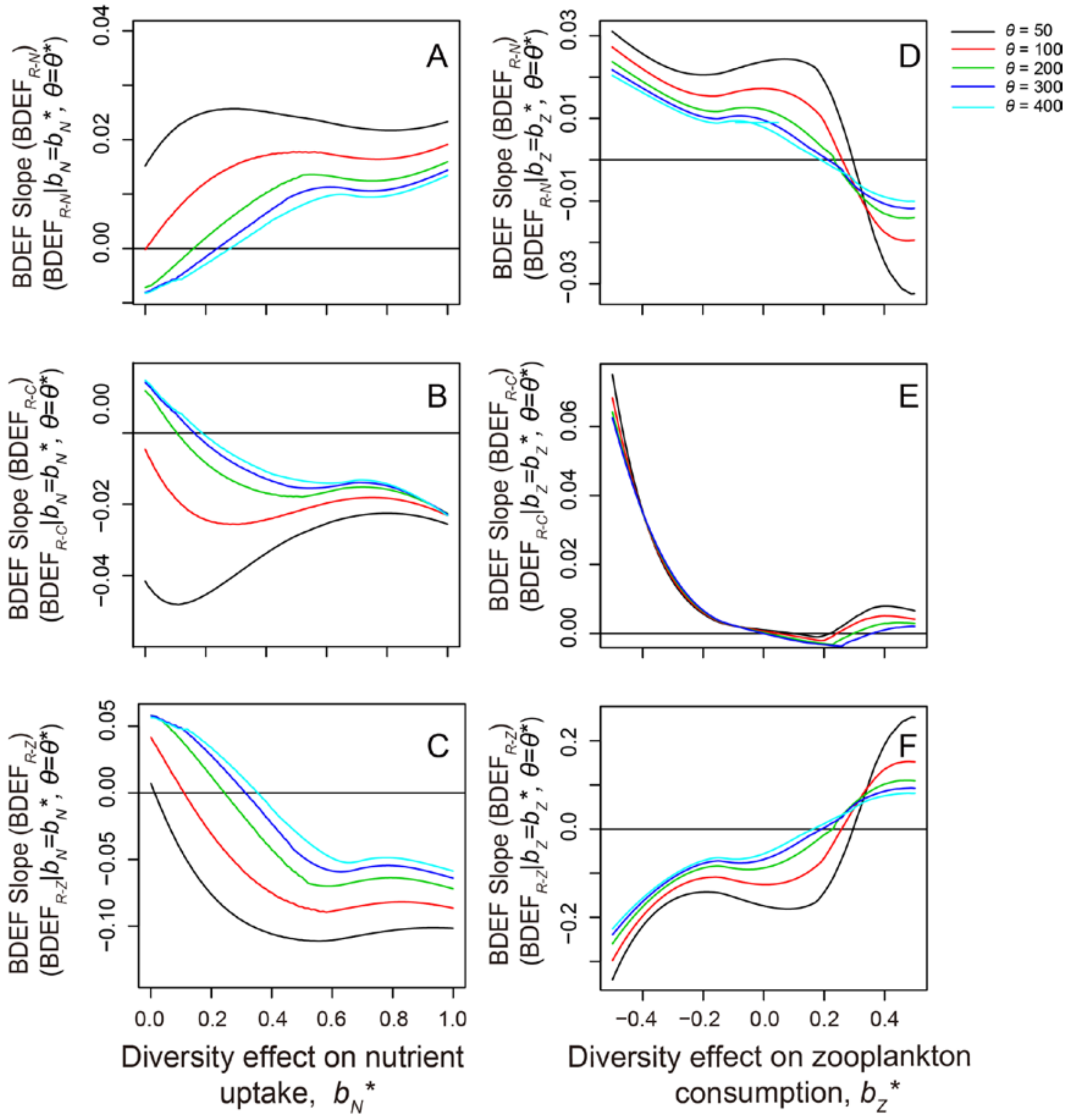

Fig. S6. The quantification of conventional BDEF relationships with slopes determined by the linear regression of average nutrient ($BDEF_{R\text{-}N}$ in *A* and *D*), phytoplankton biomass ($BDEF_{R\text{-}C}$ in *B* and *E*), and zooplankton biomass ($BDEF_{R\text{-}Z}$ in *C* and *F*) against average phytoplankton species richness. Compared with the similar conditional calculations shown in Fig. S4, we demonstrate how these BDEF relationships changes with $b_N$, $b_Z$, and $\theta$ under some specific parameter sets, ($b_N$=$b_N$*, -0.5≦$b_Z$≦0.5; *A*-*C*) and (0≦$b_N$≦1.0, $b_Z$=$b_Z$*; *D*-*F*) with various $\theta$=$\theta$*. Firstly, we show how the BEF slopes vary with respect to $b_N$ (*A*, *B*, and *C*). With increasing diversity

effects on nutrient uptake $b_N$, (*A*) BDEF$_{R\text{-}N}$ becomes more positive; in contrast, (*B*) BDEF$_{R\text{-}C}$ is more positive under small $b_N$. This unexpected result is caused by strong top-down control of zooplankton. Specifically, increasing nutrient consumption efficiency though instantly benefits phytoplankton, the overgrowth of phytoplankton will soon be consumed by zooplankton which continuously suppresses phytoplankton biomass to a low level in most of the time (Fig. S2*A*). Interestingly, the top-down control of zooplankton on phytoplankton could be attenuated when nutrient feedback on diversity is strong (green and blue lines), which results in more positive BDEF$_{R\text{-}C}$ (*B*) and BDEF$_{R\text{-}Z}$ (*C*) under small $b_N$. Therefore, the strength of diversity effects (e.g., $b_N$) cannot always be intuitively inferred from the BEF relationship between average diversity and ecosystem functioning in nonlinear dynamical systems. Secondly, we demonstrate how the BDEF slopes vary with respect to $b_Z$ (*D*, *E*, and *F*). Our analysis revealed that more negative diversity effects on zooplankton ingestion (i.e., defence mechanism dominates) usually lead to more positive BDEF$_{R\text{-}N}$ and BDEF$_{R\text{-}C}$ (*D* and *E*) but more negative BDEF$_{R\text{-}Z}$ (*F*). Moreover, the negative impacts of phytoplankton diversity on zooplankton could be undermined under strong nutrient feedback on diversity (large $\theta$ but negative $b_Z$ in *F*). In conclusion, the existence of nutrient-diversity feedback generally makes the less significant BDEF relationships (smaller absolute values of slopes) and let the BDEF slopes becoming less sensitive to changes in the strength of multitrophic BDEF ($b_N$ and $b_Z$).